# サイバーセキュリティに関する研究倫理の調査と倫理指針の構築
## Investigation on Research Ethics and Building a Benchmark

稲垣 俊 *　　　Robert Ramirez *　　　島岡 政基 *　　　潟 謙一 *
Shun Inagaki　　Robert Ramirez　　Masaki Shimaoka　　Kenichi Magata

**あらまし** サイバーセキュリティにおける先進的な研究，特に攻撃的セキュリティと呼ばれる分野の研究は，しばしば倫理的な議論を必要とする．現状ではサイバーセキュリティ特有の研究倫理指針は存在せず，日本の攻撃的セキュリティ研究を躊躇させる一因となっている．本研究では，過去のトップカンファレンスに投稿された論文を機械学習および手作業で分類し，研究倫理に関するベストプラクティスを抽出した．抽出した知見からサイバーセキュリティに関する研究倫理の倫理指針を構築した．この倫理指針はグレー領域の研究を禁止された研究と誤認されないようにすることを目的としている．倫理指針を決定木の形で倫理指針を表現することにより，セキュリティ研究者は自分の研究で倫理的議論を必要とする点を効率的に確認することができる．本稿では手始めに一部のカテゴリのみ対象として倫理指針構築を試みたが，手法そのものは他分野にも応用が期待できる．

**キーワード** サイバーセキュリティ，研究倫理，トピックモデル

## 1 はじめに

先進的な研究には倫理的な課題が付随することがある．医学をはじめとする人間や社会に大きな影響を与えうる研究を行う場合，あらかじめその研究の倫理的妥当性を問う必要がある．情報通信インフラが末端まで行き届いた現代では，サイバーセキュリティの研究も社会に大きな影響を与える可能性がある．したがって，医学同様に研究が与える影響を十分に考慮しなければならない．

サイバーセキュリティ研究の分野には明らかに他者の権利を明確に侵害するもの，しないもの，そしてどちらとも言い難い倫理的にグレーな領域が存在する．明らかに問題のある研究や明らかに問題のない研究については，その研究活動の可否を判断することが容易である．よって，特に議論を要するのは倫理的にグレーな領域である．セキュリティ研究らは攻撃者のように考え，攻撃者よりも先に新たな攻撃手法を調査しなければならない．そのような倫理的にグレーな領域，すなわち攻撃的セキュリティ研究はサイバーセキュリティ研究者にとって有益なものとなるが，同時にその研究が社会に悪影響を与えないことを考慮しなければならない．

被験者実験を行うことが多い医学等の研究領域では，研究を実施する際に倫理委員会の承認を必要とすることがある．しかしながら，現状ではサイバーセキュリティを扱う倫理委員会の存在が希薄であると言わざるを得ない．その一因は，サイバーセキュリティの研究倫理に関して判断する基準が不明確なことにある．医学の分野では古くから研究倫理に関する議論が行われており，ヘルシンキ宣言やベルモントレポートのような倫理指針を参考とした研究が行われてきた．サイバーセキュリティの分野でもベルモントレポートの流れをくむメンロレポート [1] が 2012 年に登場したものの，それを用いた実践的な判断基準は不足している．

そこで本研究ではサイバーセキュリティに関連する研究倫理課題を洗い出し，研究倫理指針の素案を作成した．この倫理指針はグレー領域の研究を禁止された研究と誤認されないようにすることを目的としている．倫理指針構築には，おおよそ過去 5 年間のトップカンファレンス論文のベストプラクティスを参考にした．トップカンファレンスに採択されるような研究は先進的な問題を扱っていることが多く，それらは同時に倫理的な問題を扱っていることを意味する．加えて，そのような論文は倫理的な議論が十分になされていることが多く，それらは倫理指針構築に役立てることができた．本研究では 962 本の論文を対象とした機械学習のトピックモデル作成を通して，論文を 5 カテゴリに分類した．そのうち，試験的に 2 カテゴリのみを対象として倫理指針を構築したが，本手法は残る 3 カテゴリにも適用可能な汎用性の高いものとなっている．倫理指針は決定木の形をとっており，研

---

* セコム株式会社 IS 研究所, SECOM CO., LTD., Intelligent Systems Laboratory

究活動において倫理的に考慮すべき点を各段階ごとに明確に示している．

提案手法である決定木の構築によって一定の網羅性を実現する一方で，ベストプラクティスの集積とのハイブリッドなアプローチには議論が不十分な領域が残っていることを浮き彫りにした．ベストプラクティスにしても更なる議論の成熟が期待されるものもあり，現時点での倫理指針は完成形ではなく，あくまでも手法の有用性を示す一例であることを強調しておく．

## 2 先行研究

研究倫理の綱領や基準は被験者実験を扱う医学や社会学の分野で発展した．多くの倫理指針が人間の尊厳を守ることに議論の重点を置いているのは，その歴史的な背景と関係がある．倫理指針の始まりは 1947 年のニュルンベルク綱領まで遡る．ここでは被験者の自発的な同意やリスクと利益の解析，後に影響のない範囲で実験を取り下げる権利を掲げている．1964 年，フィンランドのヘルシンキ宣言ではニュルンベルク綱領に加え，被験者の利益が公益よりも優先されるべきことや，すべての被験者が十分な扱いを受けることが追加されたことで，より柔軟かつ丁寧な被験者実験が行われるようになった．その後 1974 年にアメリカで成立した国家研究法では倫理委員会の設置を義務付けた．そして，その数年後にベルモントレポートと呼ばれる研究倫理のガイドラインを発表した．ベルモントレポートでは「人格の尊重」，「恩恵」，「正義」を軸にして研究倫理のあるべき姿を描写している．これらは医学の臨床の現場では「インフォームドコンセント」や「危険性と利益の解析」，「被験者の構成な選抜」という形をとっている．

以上のような医学における研究倫理は被験者実験に重きを置いているものの，サイバーセキュリティをはじめとした ICT（Information Communication Technology）に関する研究を実施する際にも重視すべき内容を多分に含んでいる．実際，ACM や IEEE が掲げる倫理綱領でも他社に害をなさないことや社会に貢献すること，誠実であることなどを会員に求めている．これらは抽象的な概念ではあるが，医学の現場で形作られた倫理指針と同じ目的を持つことがわかる．

ICT に関するより包括的な倫理指針として Ethical Decision-Making and Internet Research やメンロレポート [1] がある．これらは先に議論された倫理指針を ICT の観点から見直し，必要な項目を付け加え，修正している．ICT の研究は医学のそれに比べ，考慮すべき関係者の数や種類が増加する．研究対象が人間のみならず，システムとなる場合もあるため，それらに対する影響も併せて考慮する必要がある．また，研究対象と相互接続するシステムやその提供者のように直接的な研究対象でないにも関わらず，影響が及ぶ関係者らも存在する．このように研究の責任範囲や影響が医学の研究とは大きく異なるため，ICT 特有の倫理指針は度々議論されてきた．2002 年，Association of Internet Researchers のメンバーは Ethical Decision-Making and Internet Research を発表した．これはインターネット研究に関する研究倫理を説いたものであり，2012 年に新版が登場した [2]．進化する技術や多様な背景に対して柔軟性を維持することを目指しているため，具体的な議論を避けた汎用的な倫理ガイドラインとなっている．2012 年に登場したメンロレポートはベルモントレポートの流れを色濃く汲んでいる．ベルモントレポートの中核をなしていた「人格の尊重」「恩恵」「正義」に加え，メンロレポートでは「法と公益の尊重」を柱としている．そのうち，メンロレポートでは倫理的な問題に関係する事例を多数掲載しているものの，その倫理的な判断は読者らにゆだねられている．

本研究では過去のサイバーセキュリティに関する研究事例を参考にし倫理的な問題を洗い出しただけでなく，研究活動を行うどの時点でどのような倫理的な問題が発生しうるのかを明らかにした．倫理的な判断は決定木の形で表現されている．研究者らが行おうとしている研究活動に対して特定の手法を用いようとしたとき，彼らは決定木を辿ることで当該手法が倫理的な議論を必要とするか否かを判断できる．

研究の段階ごとに倫理的な選択肢を提示することで，先行研究に比べ具体的な倫理の議論を行うことが可能となる．IRB[1] の審査官がサイバーセキュリティの専門家でなくとも，研究の際に注意すべき点を確認できるため，サイバーセキュリティにおける倫理的な議論が欧米諸国に比べ遅れている日本では特に有用な事例となるだろう．

## 3 提案手法

本研究では過去のトップカンファレンスの論文を参考にして，サイバーセキュリティに関する研究倫理指針を構築した．トップカンファレンスに採択されるような研究は先進的な問題を扱っていることが多い．それらは同時に倫理的な問題を扱う機会が多いことを意味する．そのような論文は倫理的な議論が十分になされていることが多く，それらは倫理指針構築に役立てることができた．

サイバーセキュリティ研究の中では分野ごとに，倫理的な問題が生じることがある．そこでまずは様々な論文を紐解き，その研究が関連している研究課題を明らかにする必要がある．提案手法では，まず論文の収集を行った．その後，それらを対象に（a）研究分野を表すトピックモデルの構築と（b）グレー領域/倫理の議論を含む論文集合の構築を実施した．（b）で得られた論文集合を（a）

---

[1] Institutional/Internal Review Board

表 1: 論文収集元

| カンファレンス | 年代 | 収集した論文数 |
| --- | --- | --- |
| USENIX Sec.[2] | 2013-2016 | 249 |
| IEEE S & P[3] | 2013-2017 | 183 |
| ACM CCS[4] | 2016 | 138 |
| SOUPS[5] | 2014-2016 | 65 |
| USESEC and NDSS[6] | 2013-2016 | 253 |
| CREDS[7] | 2013-2014 | 8 |
| PETS[8] | 2015-2017 | 93 |
| SSRN[9] | Any | 32 |
| All | Any | 994 |

で得られたモデルに適用すると各分野ごとに注目すべき論文集合が得られた．それらを精読し，倫理的な問題やその解決法を洗い出したのち，倫理指針を構築した．

### 3.1 論文の収集

トップカンファレンスの論文収集手法について記述する．サイバーセキュリティに関する倫理は時代や技術と共に変化する．最新の技術とそれに関する議論のみを倫理指針に反映させるためには，それに応じた論文を収集しなくてはならない．筆者らは 2012 年に登場したメンロレポートがサイバーセキュリティの研究倫理に特に影響を与えていると考え，その出版年をひとつの区切りとして論文収集を行うこととした．

収集対象となるカンファレンスは攻撃技術を含むサイバーセキュリティに関する投稿が多いものや倫理の議論を多く含むものを中心に扱った．また，今回の実験では再実験や検証を容易にするために本研究遂行時点（2017年 8 月）でオープンアクセス可能な論文のみを扱うこととした．

### 3.2 トピックモデルを用いた論文のカテゴライズ

収集した論文をそれが扱っている研究課題ごとに分類した．論文の中には論文著者らによってキーワードが設定されているものも存在するが，本手法ではそれらを用いない．自然言語処理を用いて，論文中に現れる単語やフレーズから論文をカテゴリ分けする．これには，ふたつの狙いが存在する．ひとつは本文から構成したトピックを利用することで論文著者らが設定した観点とは別の観点から論文を分類できる可能性があるということである．もうひとつは収集した論文集合の中から共通する研究課題を抽出することで，それらが抱える倫理的な問題が明確になるということである．

---

[2] USENIX Security Symposium
[3] IEEE Symposium on Security and Privacy
[4] The ACM Conference on Computer and Communications Security
[5] Symposium On Usable Privacy and Security
[6] Network and Distributed System Security, Workshop on Usable Security
[7] Cyber-security Research Ethics Dialog & Strategy Workshop
[8] The annual Privacy Enhancing Technologies Symposium
[9] Social Science Research Network

提案手法ではトピックモデルを用いて研究課題を抽出し，各論文を分類した．トピックモデルは，与えられた文章がどのようなトピックについて言及しているのかを抽出する教師無し学習の手法である．本研究では，Python実装 [3] による Latent Dirichlet Allocation (LDA) を用いた．このモデルでは，文章中に現れる単語がトピックと呼ばれる未知の事前分布に従い，各文書がトピックに関する未知の事前分布を持つことを仮定している．LDAでは単語とトピックの結合分布学習を目的としている．すなわち，データセットに現れるすべての単語，およびそれぞれの文章に現れる単語の分布を LDA に与えることで，それぞれのトピックにある単語が属する確率とそれに基づくトピック毎の単語の集合が得られる．

サイバーセキュリティに関する表現の違いにより，同様のトピックが別のグループに分類されることがある [4]．したがって，完成したモデルは手作業で確認する必要がある．提案手法では，それぞれのトピックを最もよく表現する 30 語を抜き出し，それを用いてトピック間の類似度を手作業で測った．そして，複数の類似するトピックから構成される集合をカテゴリとして扱うこととした．

トピックモデル構成の前処理として論文の PDF から単語の抽出を行い，可読不能な文字やストップワードを除外した．併せて，単語のみならず複数の単語から成る句も解析対象とするために，$n$-gram ($n = 1..5$) を用いた．

### 3.3 グレー領域の論文の抽出，倫理に言及する論文の抽出

収集した論文の中には倫理的に議論の余地があるグレー領域の論文や，すでに倫理に関する議論を行っている論文が存在した．それらは優先的に調査すべき論文であるため，集めた論文からグレー領域[10] に関する論文や倫理に言及している論文を抽出し，それぞれの論文集合を作成した．

グレー領域に関する論文か否かはタイトルと概要から判断した．ここでは理論研究でないもの，研究者によって管理された環境外に影響を及ぼす可能性のある研究，そして倫理的に不明確な点を含む研究をグレー領域の論文とした．加えて，被験者実験を含む研究もグレー領域研究とした．

明に倫理に関して議論している論文集合を作成するために，正規表現による文書検索を用いた．以下のキーワード `ethic*`, `moral*`, `IRB or REB`[11] のうち，いずれかが論文中に現れた場合，それを倫理的な議論を含む論文として扱った．

---

[10] より正確には明らかに白とは言い切れない領域の論文．
[11] Research Ethics Board

### 3.4 倫理指針の構築

グレー領域または倫理に関する議論を含む論文を，3.2 節で作成したカテゴリごとに分類した．そして注目するカテゴリに属する論文を精読することで，そこから倫理的な問題点や倫理的な議論を抽出した．提案手法では，収集した倫理的な問題や議論を系統立て決定木の形で表現した．

根から葉までのノードが研究活動を示し，葉はそれに対する倫理的な判断を示す．葉は次の 4 種類，

- 禁止 (Prohibits): 当該活動は非倫理的である
- 許可 (Permits): 当該活動は非倫理的でない
- 必要 (Demands): 当該活動を実施しないことが非倫理的である
- 要検討 (TBD): 現時点では判断できない

から構成される．禁止，許可ノードは当該活動の倫理性についての判断例である．なお，ここで許可ノードは必ずしも常に受容可能を意味するわけではないことに気を付けなければならない．たとえば，本倫理指針では脆弱性の公開について，それが既にエクスプロイトされているものであれば，公開を許可するとした．しかしながら，公知の脆弱性でもその情報を改めて世に広めることで被害が拡大する可能性がある．したがって，研究者らは本倫理指針の示す項目にのみ依存するのでなく，自らが置かれた状況を踏まえて倫理的な判断を下す必要がある．必要ノードはグレー領域の研究に付随する対策のベストプラクティスを示している．これらは特に重要なものであり，論文中でも倫理的な考察を与える必要がある．要検討ノードはベストプラクティスを見つけられなかった，または今後の議論を要する部分である．今後も様々なベストプラクティスを参考にしつつ，本倫理指針の拡充に役立てることを期待する．

中間のノードにも特殊な種類のものが存在する．"XOR" ノードはそれ以下に互いに相いれない選択肢を複数持ちうるノードである．このノードが登場した場合，倫理指針の利用者は XOR ノード以下に接続されたノードのうちのいずれかを選択しなければならない．また，それ以下の葉ノードに到達するための条件を現時点では決定しきれない場合，"condition" ノードを仮に設置した．これは葉の要検討ノードとほぼ同義である．

## 4 実験結果

3 章で説明した手法を用いて倫理指針の構築を行った．表 1 は今回収集した論文の収集元と対象とした年代，収集した論文の本数を示している．

これらの論文を基にトピックモデルを構築した．ここで，SSRN はカンファレンスではないので除外した．トピックの数は 50 を指定し，他のパラメータはトピックモデルライブラリ Gensim[3] のデフォルト値を用いた．得られたトピックに属するフレーズを手作業で 13 のトピックカテゴリにまとめた．すなわち，あるトピックカテゴリは複数のトピックから構成されている．表 2 はトピックカテゴリとそれに属するトピックおよびトピックに属する句の代表例を表している．

次に，収集した論文集合からグレー領域に属する論文と倫理に言及している論文を抽出した．前者は 234 件であり，後者は 200 件であった．どちらにも含まれている論文は 49 件であった．グレー論文と倫理に言及している論文の和集合を先に構成したトピックならびにトピックカテゴリに分類した．分類の基準は次の 2 つ

1. ある論文とトピック間の尤度が最も高いトピック
2. ある論文とトピック間の尤度が 2 番目高く，その値が 0.1 以上となるトピック

表 2: それぞれのトピックカテゴリに属するトピックとそれを構成する代表的な句

| カテゴリ | トピック | 上位の単語 |
|---|---|---|
| authentication | 7 | native code, collected data, legitimate users, tenth symposium, shoulder surng, session key, users may |
| security behavior | 4 | previous studies, security behaviors, information security, identity theft |
| low level/OS/IoT | 6 | sa fed, linux kernel, control flow, power consumption, address space, operating system |
| privacy controls | 5 | privacy notices, privacy settings, privacy concerns, users privacy, data practices |
| PII collection | 6 | internet users, statistically significant, demographic information, personal information, one participant, phone number |
| vulnerabilities | 3 | et al, security symposium, security policy |
| encryption | 4 | computer science, security actions, malformed blocks, public key |
| user oriented design | 3 | participants reported, security questions, would like |
| identity management | 1 | mobile phone |
| online measurements | 7 | touch id, per day, ip address, data collection, mturk workers, social networks, private key |
| website analysis | 2 | new york, available http |
| version control | 1 | ca usa |
| bank account | 1 | amazon com |

である．ここで2番目の基準の閾値はヒューリスティックに決定したが，分類結果のうち10本の論文を読み，その分類が妥当であることを手作業で確認した．また，この段階でもSSRNなどカンファレンスから収集していない文章は分類から除外した．表3はその分類結果を示している．

本研究では，"vulnerabilities"と"online measurements"に絞って論文を精読することとした．Vulnerabilitiesには15本の論文が含まれており，これらをすべて精読した．Online measurementsには61本の論文が含まれていたが，そのうちグレー領域の論文かつ倫理に言及している論文である13本を精読した．そして，これらの28本に含まれる倫理的な課題やそれに対する問題解決法を系統立ててまとめた．

### 4.1 決定木を用いた倫理指針の表現

先に挙げた28本の論文から得られた倫理的な問題や解決法，および参考として収集した他の論文を基に手作業で倫理指針を構築した．構築した倫理指針は5つの倫理指針メインクラスと複数のサブクラスから構成される．表4はメインクラスとその直下のサブクラスを示している．また，図1は決定木の一部を示している．根から葉までのノードがある研究活動を示し，葉はそれに対する倫理的な判断を示す．倫理的な判断はトップカンファレンスから得られたものや論文著者らの議論によって決定した．したがって，筆者らはそれらが必ずしも正しい判断であることを主張しない．さらに決定木構築時に，決定できなかった倫理的な判断も多数存在した．これは本倫理指針を利用する研究者が所属する組織のポリシーが倫理的な判断に関与する場合があるためである．

以下では決定木をメインクラスごとに構成する際に用いた論文とそこから得られた知見について記す．

#### 4.1.1 Software Examination

脆弱性の研究それ自体は関係者の同意の下で許容可能である．同意が得られない場合には，ライセンスや倫理指針の許容範囲内で研究を実施しなければならない．たとえば，多くの論文ではオープンソースのシステムに対して脆弱性の研究が実施されている [5, 6, 7, 8, 9, 10]．

脆弱性の研究は必ずしもリバースエンジニアリングを必要としない．しかしながら，それを必要とするときには契約または知的財産権の侵害を起こさないよう気を付ける必要がある．

マルウェアは脆弱性を用いて攻撃を展開しうるため，これは本項目で扱う．マルウェアに関する研究を行う際には，それが与える影響を十分に考慮し，悪影響を低減しなくてはならない．対象の同意が得られ，本倫理指針の他の項目を遵守する場合，本倫理指針ではコンピュータをマルウェアに感染させる動的解析に関しても，それが外部に害を及ぼさない限り受け入れられる．たとえば，まずローカルの隔離された環境でマルウェアを実行し，その影響を把握するべきである．

マルウェアの送信に関する倫理的な側面は様々な要因に関係する．たとえばマルウェアの作成者，送信先，送信してからそのままにしておく時間，本件の関係者，そして送信目的である．また，これらはマルウェアの定義によっても変化しうる可能性がある．調査目的でAppleストアにマルウェアを送信した研究が存在する [11]．この研究者はマルウェアをAppleストアから自身の端末にダウンロードしたのち，それをAppleストアから直ちに削除している．また，Appleの統計を用いて，同研究者ら以外がマルウェアをダウンロードしていないことを確認している [12]．

脆弱性の公開は関係者らに影響を与えることがある．多くの論文では脆弱性の公開に先立ち，ベンダや関係者に連絡している [6, 7, 11, 12, 13, 14, 15, 16]．その際，脆弱性が新しいか，誰が脆弱性を開示するか，脆弱性の開示範囲，脆弱性開示のタイミング，開示の取り扱い，そして関係者との協力等に基づき，情報開示に関する倫理的な判断を下す必要がある．

#### 4.1.2 Privacy

本倫理指針ではPII(Personal Identifiable Information)とその他の情報を区別していない．これは正確なPIIの定義が困難なためである．その代わり，本研究では人や動物，システムに関して収集したすべてのデータを対象とし，すべて等しく慎重に扱うよう求めている．

また，そのようなデータを扱う際には，あらかじめデータ取得時から廃棄までの流れを規定し，それに沿ってデータを管理する必要がある．このデータパイプラインはセキュアに構築され，再利用可能なものとして規定することが望ましい．

著者らが収集した論文のなかには機微なデータを扱う際に，必要最低限のデータのみを収集し，それに暗号化や

表3: グレー領域の論文および倫理に言及している論文をトピックモデルに適用した結果

| カテゴリ名 | 論文数 |
|---|---|
| vulnerabilities | 15 |
| online measurements | 61 |
| low level/OS/IoT | 20 |
| security behavior | 40 |
| personal information collection | 61 |

---
[12]当該論文は日本国外で発表された論文である．日本で研究する際には関連法に該当しない様な配慮が別途必要である．

個人識別子の削除を施していたものもあった [17, 18, 19]．データの保管のみならず，送信時にも安全な送信方法のみを利用することを記している論文も存在する [18]．いくつかの論文では，実験の後に収集したデータを破棄していることを明示している [5, 19]．

### 4.1.3 Autonomy

本項目では，第三者の持つシステムや人に対する干渉に言及している．

同意を得たのちに第三者のシステムにアクセスすることは受け入れられるが，研究によって与える影響は考慮しなければならない．アクセスの影響が測れず，同意なくシステムにアクセスするのであれば，それは非倫理的である．たとえば，ボットネットの操作に関する研究論文が存在する [20]．当該論文の中で著者は感染したボットに未承認のコミュニケーションを用いることで攻撃ができ，脆弱性を利用することで時にボットの感染を止めることができると述べている．しかしながら，その一方でその手法は付帯的損害を発生させるため，非倫理的であるとしている．別のボットネットに対する研究 [5] において，彼らは意図的に被害を出すことを非倫理的であるとしている．しかしながら，我々の倫理指針の中では何もしないことを非倫理的としている．すなわち，影響がないことを想定するだけでなく，影響の度合いを評価しなければならない．

自動的にデータを収集する際にも配慮が必要である．ある研究では，Web サイトから大量にデータを収集することを目的としているが，それぞれのサイトでは過度のアクセスを実施していない [14]．同様に Web サイトのスクレイピングに関しては，既存のデータベースを利用することでサーバに対する新たな負荷をかけずにデータを収集する研究も存在する．

### 4.1.4 Human and Animal Subjects

本倫理指針では人や動物に対する被験者実験を慎重に実施することを求めている．

被験者実験の一部では事前に同意を得られないものも存在する．そのような場合，メンロレポートにも記載されていたように実験後に十分な説明を行わなくてはならない．たとえば，ある研究ではソーシャルネットワーク上で被験者を騙したものの，検証の終わりにそれを明らかにした [21]．また別の研究ではコミュニティサイトである Craigslist に偽広告を掲載し，それに対するスパム収集を実施した [18]．この研究では，同サイトの正規利用者が誤って当該広告に連絡する可能性がある．このとき，著者らは研究目的を明かし，適切な対応を取る手続

表 4: 倫理指針のメインクラスとその直下のサブクラス

| メインクラス | メインクラス直下のサブクラス |
| --- | --- |
| Software Examination<br>（他人に作成されたプログラムの理解に関する項目） | Vulnerability Research<br>Reverse Engineering<br>Malware<br>Disclosure |
| Privacy<br>（第三者の人またはシステムに関する情報管理に関する項目） | Collecting Data<br>Handling Data<br>Publishing Data<br>Transferring Data To Third Parties |
| Autonomy<br>（第三者のシステムに作用する項目） | Web scraping<br>Accessing others' systems |
| Human and Animal Subjects Testing<br>（被験者実験に関する項目） | Deceiving human or animal test subjects<br>Misleading, false, or deceptive advertising<br>Honeypots<br>Criminal and Unethical Services<br>Consulting with REB or IRB |
| General Rules<br>（指針自体に関する項目や全体的に適用可能な項目） | Terms of Service<br>Ethical consistency<br>Documentation and Accountability<br>Definitions |

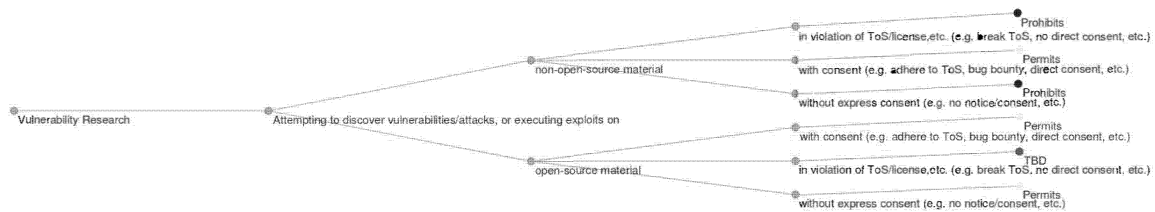

図 1: 決定木で表現した倫理指針の例

きを用意していたことが記されていた.

本倫理指針ではハニーポットの利用を許容するが, それが他者に悪影響を与えるようであれば, その利用を禁ずる [14]. ハニーポットは通常, 同意や後追いの説明を行わずに情報を収集するツールである. しかしながら, ハニーポットが与える影響を最小限に留めなくてはならない. 先に紹介した偽広告に関する研究 [18] では正規のユーザには魅力的でない広告を出すことで, 一般ユーザが偽広告に惹かれる可能性を低減した. その結果, 正規ユーザからのアクセスはなかったと当該論文の著者らは記している.

すべてのサイバーセキュリティプロジェクトにおいて, 研究者らは IRB をはじめとする倫理審査委員会に助言を求める必要がある [9, 16, 19, 22]. ここで許可されたものが倫理指針における許可ノードに該当する. また, 医学や様々な大学の倫理指針を参考にし, 被験者実験に関する項目も設けた. すなわち, 人や動物が研究に係るとき, その確証が持てない場合でも審査委員会に助言を求めなければならない. このようなルールは本倫理指針を用いる研究者以外も準拠すべき項目である.

### 4.1.5 General Rules

本項目はサイバーセキュリティ研究において広く適用できる事柄である. 本節で示した事柄を守り研究を遂行するためには, 適切な研究プロセスおよびレビュープロセスを構築しなければならない. ここで特に重要となるのは研究に関する倫理的な管理者の存在である. 彼らは研究者らから倫理的な相談を受けたり, 研究の可否に関する最終判断を下さなければならない. 同時に, 研究者は相談すべき相手として研究管理者の存在を把握しておく必要がある.

サイバーセキュリティ研究者らは自らが実施する研究について, それが外部に対して何らかの悪影響を与える可能性を常に考慮しなければならない [23]. また, 外部に対する悪影響の発生が予想できるときには, それらを十分に緩和する必要がある.

セカンドオピニオンは判断を誤った際のセーフガードとなりうる. 研究者らは何かを相談する際, 倫理的な管理者のみに相談するのではなく, 他の研究者に倫理的な疑問点を相談することも重要である. 倫理的な管理者は本研究倫理指針を適切に活用するためになくてはならない存在である. 研究者の所属組織に倫理委員会が存在しない場合には, それに代わる担当者を据えるべきである. また, 研究プロセスと責任の明確化のために文書化を徹底することも必要である [1, 23].

様々な項目で同意について言及した. 一般的な利用許諾は契約であるため, それを反故にすることは禁止されている. しかしながら, もし利用許諾が本倫理指針の他項目から乖離しており, 倫理的でないことが認められる場合, 利用許諾の準拠を理由に倫理的な正当性を主張することは禁止する. 例えば, Facebook の利用許諾を盾に, ユーザが同意すればその友人の顔画像を顔認識の研究に使用してもよいと主張したとしても, 本倫理指針の解釈では, これは倫理的ではないとした.

## 5 考察

本章では先に提示した倫理指針の導出手法や倫理指針の活用に関する考察を行う.

### 5.1 本研究の制限

本倫理指針の構築に直接用いた論文は 2013 年以降の 28 本の論文のみである. 収集した論文の多くはトピックモデルを構成するためだけに利用したものである. これらを基にした倫理指針がサイバーセキュリティの研究領域を十分にカバーしていることを筆者は保証できない. また, サイバーセキュリティ研究に関するすべての倫理的項目を提示しているということを保証できない. しかしながら, これらの論文からも広く受け入れられる実践的な知見を集約することができた. 今回対象とした論文はすべて英語の論文のみであるが, その出自はヨーロッパ, 北アメリカ, アジア, 中東を含む 11 か国と幅広いものであった. したがって, 倫理観が特定の国または地域に偏ることは回避できたと考えている. 一方それらの論文を精読し, 倫理に関する項目を抽出したのは母国語が英語である者一名である. 構築した倫理指針はその者によるバイアスがかかっている可能性があり, それを緩和するために, 筆者ら以外の識者と議論を行った.

提案手法を基に本倫理指針の内容を厚くするのであれば, 最新のトップカンファレンスの論文を用いてトピックモデルを構成し続ければよいだろう. しかしながら, 技術や時代の変化に伴い倫理感が変化する可能性がある. そのような場合には, 変化の度合いを見定め, ルールの一部を適切に変更する必要がある.

### 5.2 今後の課題

先にも記した通り, 本倫理指針は未だ発展途上である. 本節では, この倫理指針をより向上させるための今後の課題について議論する.

一番に挙げられるのは, 本倫理指針の内容の充実である. 今回精読しなかった領域の論文から倫理に関する議論を取り上げ, それらの領域が持つ倫理的な問題を洗い出すことで, より広範な倫理指針を構築することができるようになるだろう. また, より新しい論文を集め続け, それらの知見を本倫理指針に適用してもよいだろう.

先に本倫理指針の強制力を増すことで, 研究者らにこれを利用させる手法を説いた. しかしながら, 倫理指針

をより便利なものにすることで研究者らに自発的に倫理指針を利用してもらう方法もある．たとえば，本研究で得られた決定木をインタラクティブに実装し，研究者らが行おうとしている研究が倫理的であるかどうかを確認できるようにしてもよいだろう．

本倫理指針ではデータ処理手法をあらかじめ定め，その手法を再利用にすべきであると説いた．そのようなデータパイプラインの例を提示することで，本倫理指針の適用が容易になる可能性がある．

## 6 おわりに

本研究では過去のサイバーセキュリティに関する研究事例を参考にし，倫理的な問題を洗い出す手法を提案した．倫理指針は決定木として表現されており，セキュリティ研究者は自分の研究で倫理的議論を必要とする点を効率的に確認することができる．また，IRB の審査官がサイバーセキュリティの専門家でなくとも研究の際に注意すべき点を確認できるため，サイバーセキュリティ研究倫理の入口として本倫理指針を用いることもできる．

## 参考文献